\documentclass[aps,amsfonts,pra,showpacs,nobibnotes,nofootinbib,tightenlines]{revtex4}
\usepackage{amsmath}
\usepackage{amssymb}
\usepackage{graphicx}
\begin{document}
\def\bra#1{\langle #1 |}
\def\ket#1{| #1 \rangle}
\newcommand{\p}{\partial}
\newcommand{\llll}{ \ell (\ell + 1)}
\newcommand{\kk}{{k_0}^2-k^2}
\title{Implications of the Present Bound on the Width of the $\mathbf{\Theta^{+}(1540)}$
}\author{R.~L.~Jaffe}\email{jaffe@mit.edu}
\author{Ambar Jain}\email{ambarj@mit.edu}
\affiliation{Center for Theoretical Physics \\
Department of Physics and Laboratory for Nuclear Science\\
Massachusetts Institute
  of Technology \\ Cambridge, MA 02139, USA}
\begin{abstract} 
	The recently reported $\Theta^{+}(1540)$ exotic baryon
	seems to be very narrow: $\Gamma(\Theta)<$ 1 MeV according to some
	analyses.  Using methods of low energy scattering theory, we
	develop expectations for the width of the $\Theta$, an elastic
	resonance in $KN$ scattering in a theory
	where the characteristic range of interactions is $\sim$ 1 Fermi. 
	If the $\Theta$ is a potential scattering resonance, generated by
	the forces in the $KN$ channel, its width is hard to account for
	unless the $KN$-channel orbital angular momentum is two
	or greater.  If the $\Theta$ is generated by dynamics in a confined channel, its coupling to the
	scattering channel is at least an order of magnitude less than
	the coupling of the $\Lambda(1520)$ unless $\ell\ge 2$.  Either
	way, if the $\Theta$ proves to be in the $KN$ $s$- or $p$-wave,
	new physics must be responsible for its narrow width.
\end{abstract}
\vspace*{-\bigskipamount}
\preprint{MIT-CTP-3523} 
\maketitle 
\section*{ 1.~Introduction}

Reports of a narrow, light exotic baryon, the $\Theta^{+}(1540)$, have
stirred considerable interest over the past year.  Although the
existence of the $\Theta^{+}(1540)$ has yet to be decisively
confirmed\cite{trilling}, it certainly must be taken seriously. 
Perhaps the most striking characteristic of the $\Theta$ is its narrow
width.  Most sightings quote only upper limits on $\Gamma(\Theta)$
determined by experimental resolution.  These limits are as small
as 9 MeV\cite{Nakano:2004cr}.  Cahn and Trilling have extracted
$\Gamma(\Theta)=0.9\pm 0.3$ MeV from their analysis of Xenon bubble chamber
data\cite{Cahn:2003wq}.  The $\Theta$ is seen as a bump in $K_{S}p$
or $K^{+}n$ invariant mass spectra.  These are the only open channels
to which it can couple, therefore $\Theta$ must appear in some
partial wave of definite angular momentum, $j$, isospin, $I$, and orbital angular momentum, $\ell$, in $KN$ scattering
The elastic scattering cross section is $\sigma^{I}_{\ell, j}(k)=2\pi(2j+1) \sin^{2}\delta^{I}_{\ell,j}
(k)/k^{2}$ for center of mass momentum $k$.  At a narrow, elastic
resonance the phase shift must rise rapidly through $\pi$.  Therefore
the $\Theta$ must appear as a classic Breit-Wigner resonance in $KN$
scattering.  Re-evaluations of old $KN$ scattering data fail to show
any evidence of the $\Theta$, leading to upper limits on its width
in the range of 0.8 -- ``a few'' MeV 
\cite{Nussinov:2003ex,Arndt:2003xz,Cahn:2003wq,Sibirtsev:2004bg,trillingagain}.  

It is clear that the $\Theta$ is narrow, but so are other hadrons
like the $\phi(1020)$ ($\Gamma(\phi)=4.26\pm 0.05$ MeV) and
$\Lambda(1520)$ ($\Gamma(\Lambda(1520))=15.6\pm 1.0$ MeV).  We think
we know what makes these states narrow, but so far we lack a good
physical explanation why the $\Theta$ should be so stable.  Its small
width poses a challenge for any theoretical interpretation.  In this
paper we use low energy scattering theory to  gauge how great is the challenge.  In particular we
compare the width of the $\Theta$ with what we should expect for a
resonance that appears at 1540 MeV in $KN$ scattering in QCD, where
the characteristic range of interactions is $\sim$ 1 Fermi.

 The analysis of low energy hadron-hadron scattering in QCD is, in general, a very difficult problem.  Typically many channels are open, inelastic processes (particle production) are important, and relativistic effects cannot be ignored.  The $\phi(1020)$, for example, is thought to be predominantly a $J=1$, $I=0$, $\bar s s$ state strongly coupled to $\bar KK$.  At 1020 MeV the kaons are non-relativistic.  However the   $\phi$ couples strongly to $\pi\pi$  which must be described relativistically, and to $\pi^{+}\pi^{-}\pi^{0}$ which involves production and possibly the $\rho$ resonance in the final state.  In general it is not possible to subject QCD models  of hadrons to definitive tests by making predictions for hadron-hadron scattering.

The $\Theta(1540)$ is a striking exception:  particle production is not kinematically possible; only one channel is open\footnote{Provided the $\Theta$ has definite isospin.}; and the motion is non-relativistic ($v^{2}/c^{2}= 0.08$ for the nucleon and 0.30 for the kaon).  Therefore it should be possible to use the methods of non-relativistic, two-body scattering theory, based on the Schr\"odinger equation, to help characterize this peculiar particle.  

Roughly speaking --- we shall be more precise in the following sections ---  a resonance at low energy (a) can arise from the potential in the $KN$ channel or (b) can be induced in $KN$ scattering by coupling to a closed or confined channel.  These mechanisms differ both in principle and in practice.  Case (a) is the one familiar from single channel potential scattering.  The analysis of Case (a) begins with a non-relativistic $KN$ potential, $V(r)$, which may depend on isospin, $\ell$ and $j$.  If $V(r)$ is attractive enough,  it can generate a resonance in partial waves with $\ell>0$, where the angular momentum barrier stabilizes the resonance.  A resonance in the $s$-wave requires a potential with repulsion at long distances and attraction within.  The resonance is a feature of  the $KN$ continuum.  If the interaction were ``turned off'', the resonance would become broader and broader until it disappeared into the continuum.  Case (b) is quite different.  The resonance exists \textit{ab initio}, independent of the $KN$ interaction.  In the case of a non-exotic channel in, say, $\overline K N$ scattering, it could be a $qqq$ state  which happens to have the same quantum numbers as some $\overline KN$ partial wave.  If quark pair creation is ignored, it is stable --- a zero width bound state immersed in the meson-nucleon  continuum.  As the coupling is ``turned on'', the state develops a width and appears as a resonance in meson nucleon scattering.  Cases (a) and (b) behave oppositely as the relevant coupling is sent to zero:  potential resonances disappear from the scattering channel by subsiding into the continuum, confined channel resonances disappear because their widths go to zero.

Our prejudice is that most hadrons should be regarded as closed channel resonances: they decouple from hadron-hadron scattering as quark pair creation is turned off.  However, exotics in general, and the $\Theta$ in particular, could be an exception. The $\Theta$ could be a $KN$ scattering state, since its minimal quark content is $qqqq\bar q$.  Therefore we are obliged to compare the predictions of both mechanisms, (a) and (b), with the properties of the $\Theta$.     

Case (a) --- single channel, non-relativistic potential scattering --- needs no introduction.  In contrast, Case (b), non-relativistic scattering with confined channels, is not familiar to particle physicists.  The natural formalism for discussing resonances in closed or confined channels in non-relativistic scattering was developed by Fano and Feshbach\cite{fano,fesh}.  It is well known  in atomic and nuclear physics where the resulting states are known as ``Feshbach resonances''.  The basic idea is extremely simple.  In the case of two channels, one open, the other closed/confined,  the scattering particles make a transition to the closed channel where there is assumed to be an energy eigenstate nearby.  The system sits in this state until the transition is reversed.  If the transition potential is weak, the resonance is narrow.  We explain the method further when we apply it to the $KN\Theta$ system in Section 3.

We consider both possible origins for the $\Theta$.
In Case (a) we find that the $\Theta$'s narrow width is unnatural
unless the orbital angular momentum in the $KN$ system is at least
$\ell=2$, or better $\ell=3$.  There is no resonance for $\ell=0$. 
For $\ell=1$, a resonance at 1540 MeV with width of order 1 MeV
requires a unacceptably short range or otherwise unnatural
potential\cite{jw}. Further results for this case are
presented in Fig.~1.

In Case (b) we find the strength of coupling between the scattering channel and the confined/closed channel as a function of the range, $b$, and orbital angular momentum $\ell$, required to produce a width of order 1 MeV. In order to normalize this strength, we compare it with the results of an identical calculation for the $\Lambda(1520)$.  The $\Lambda(1520)$ is an excellent candidate for comparison because it is well described as a $uds$-configuration that requires quark pair creation in order to decay into $\overline{K}N$.  Since it has approximately the same mass as the $\Theta$, the kinematics of $\Lambda(1520)\to \overline{K}N$ and $\Theta\to KN$ are nearly identical.  The $\Lambda(1520)$ partial width into $\overline{K}N$ is 7 MeV.\footnote{The one way in which the $\Lambda(1520)$ is not entirely analogous to the $\Theta$ is that it also couples to $\pi\Sigma$.  For the purposes of this estimate we ignore the $\pi\Sigma$ channel and use the $\overline{K}N$ partial width instead of the full width of the $\Lambda(1520)$.} Our conclusions in this case depend on $b$ and are summarized in Fig.~2.  In brief, for $\ell=0$ and $b\approx$ 1 fm, a state with the width of the $\Theta$ would have to couple to $KN$ with strength of order 1/50 of the $\overline{K}N\Lambda(1520)$ coupling.  For $\ell=1$ the strength would have to be of order 1/10 of the $\overline{K}N \Lambda(1520)$ coupling, and for $\ell=2$, 1/3.

Our analysis has serious consequences for models of the $\Theta$.  The important distinction between models is whether they rely on the $KN$ potential to generate the resonance or they allow for confined/closed channels.  Chiral soliton models (CSM) fall into the first class and quark models can, at least in principle, lie in the second.

The width of the $\Theta$ has been discussed in the chiral
soliton model.  Indeed a narrow width was predicted before the
particle was reported\cite{dpp,thetawidth}.  However these predictions were not based on explicit calculations of $KN$ scattering, but rather on the extraction of coupling constants using effective field theory.  
The meson-nucleon continuum in the CSM model is described by quantizing mesonic small oscillations in the soliton background\cite{Mattis:1984dh}.  Partial wave amplitudes can be projected out using methods of group theory.  Since the model includes nothing beyond mesons and baryons (which are coherent states of the meson field), it appears to fall into the category of $KN$ potential models.  This makes the extremely narrow width of the $\Theta$ a mystery in the CSM unless a) $\ell\ge2$, b) the $KN$ potential is very unusual, or c) the resonance arises through coupling to some heretofore unknown closed or confined channel.  

Quark models of the $\Theta$ fare a little better.  Correlated quark
models (diquarks, triquarks, {\it etc.\/}) introduce color
configurations of the $qqqq\bar q$ system which cannot separate into
$KN$\cite{jw1,Karliner:dt,Nussinov:2003ex}.  If so, Case (b) would
apply.  The configuration barrier that prevents $\Theta$ decay has to be quite high.  $\Lambda(1520)\to \overline{K}N$ is suppressed by quark pair creation and $\ell=2$ with a resulting partial width of 7 MeV.  The barrier preventing the $qqqq\bar q$ system from reorganizing into $KN$ would have to be almost two orders of magnitude greater than this if it has $\ell=0$ and one order of magnitude greater for $\ell=1$.  This is a significant constraint on quark models.

The remainder of this paper is organized as follows:    In Section 2 we analyze the $\Theta$ as a potential scattering resonance  (Case (a)) and in Section 3 we analyze it as a bound state in the $KN$ continuum (Case (b)).

\section*{2. One channel, non-relativistic potential scattering}

Resonances in non-relativistic potential scattering arise from the
interplay of attraction and repulsion.  There are no resonances in a
uniformly attractive potential, only virtual states or bound states. 
The angular momentum barrier, $\hbar^{2}\ell(\ell+1)/2\mu r^{2}$, adds
a repulsive term which can generate resonances for $\ell\ne 0$ in
otherwise attractive potentials.  To generate a resonance in the
$s$-wave it would be necessary to introduce a potential that is
repulsive at long distances and strongly attractive at shorter
distances.  We put such unusual potentials aside and limit ourselves
to the uniformly attractive case.  Under these conditions potential
scattering cannot explain a narrow resonance in
the $s$-wave.

For $\ell\ne 0$, the narrower the resonance at a fixed center-of-mass
momentum, $k_{0}$, the shorter the range and greater the depth of the
potential.  A shorter range produces a higher barrier but must be
accompanied by a greater depth in order to keep $k_{0}$ fixed.  The
question is, therefore, whether the width of the $\Theta^{+}(1540)$
requires a range which is unnaturally small or equivalently, a
potential that is unnaturally deep.  To answer the question
quantitatively we must choose a specific form for the potential
$V(r)$.  For our purposes the simplest form, a potential hole, or
``square well'', $V(r)=-V_{0}\theta(b-r)$, is adequate.  Smoothing out
the edges of the square well would not change our results
significantly.  Also we have checked that using a Yukawa potential
$V_{Y}(r)=-V_{0}e^{-r/b}/r$ does not change our results
qualitatively.  In fact a square well is probably more realistic than
a Yukawa because the singularity at $r=0$ in the Yukawa
potential is not realistic in QCD since the kaon and nucleon are not
point particles.

In the channel with angular momentum $\ell$, the phase shift,
$\delta_{\ell}(k)$, for the square well is given by an elementary
calculation,
\begin{equation}
\cot\delta_{\ell}(k)=\frac{k n_{\ell}'(kb) j_{\ell}(qb) - q 
n_{\ell}(kb) j_{\ell}'(qb)} {k j_{\ell}'(kb) j_{\ell}(qb) - q 
j_{\ell}(kb) j_{\ell}'(qb)},
\label{phaseshift}
\end{equation}
where $q=\sqrt{k^2 + 2\mu V_0}$ and $j_{\ell}$ and $n_{\ell}$ are
spherical Bessel and Neumann functions respectively.  The resonance,
if there is one, is associated with a pole in the scattering
amplitude, $f_{\ell}(k)$, which is related to $\delta_{\ell}(k)$ by $k
f_{\ell}(k) \equiv 1/(\cot \delta_{\ell}(k) - \imath)$.  For a narrow
resonance any of the conventional definitions of the resonance
location and width will suffice.  To be specific we use the
definitions suggested by Sakurai \cite{sakurai}, according to
which the resonant energy $k_0^2/2 \mu$ is defined by $\cot
\delta_{\ell}(k_0) =0$ and width by
\begin{equation}
\Gamma = - \frac{2 k_0}{\mu} \biggl [ \frac{d \cot\delta_{\ell}(k)}{dk} 
\biggr \vert_{k=k_0} \biggr ]^{-1}.
\end{equation}
 This   is meaningful only 
if $\Gamma/2 \ll k_0^2/2\mu$. Of course we are interested in 
generating a narrow resonance, therefore in presenting the results 
below we restrict to values of $b$ for which this strong inequality 
is 
satisfied.

Our results are shown in { Fig.~1(a-c)} for $\ell=1, 2$, and $3$, where
we plot the width and the depth of the potential as functions of the
range, $b$, the depth having been adjusted to keep the resonance at
$k_{0}=270$ MeV corresponding to $M(\Theta)=1540$ MeV. In Fig.~2(d) we
show the same information for a Yukawa potential for $\ell=1$.  We
conclude that a narrow $\Theta$ at 1540 MeV cannot be obtained from a
``typical'' potential for $\ell=0$ or $\ell=1$.  Even $\ell=2$
requires quite a short range $b\approx 0.24$ fm for a width of 
1 MeV.

Should the $\Theta$ be established to occur in the $s$- or $p$- wave, 
its origin cannot lie in the forces between kaons and nucleons 
unless they are characterized by a distance scale very different 
from $1/\Lambda_{\rm QCD}$ or involve some exotic behavior like 
repulsion at long distances.

\begin{figure}
\includegraphics[width=18cm]{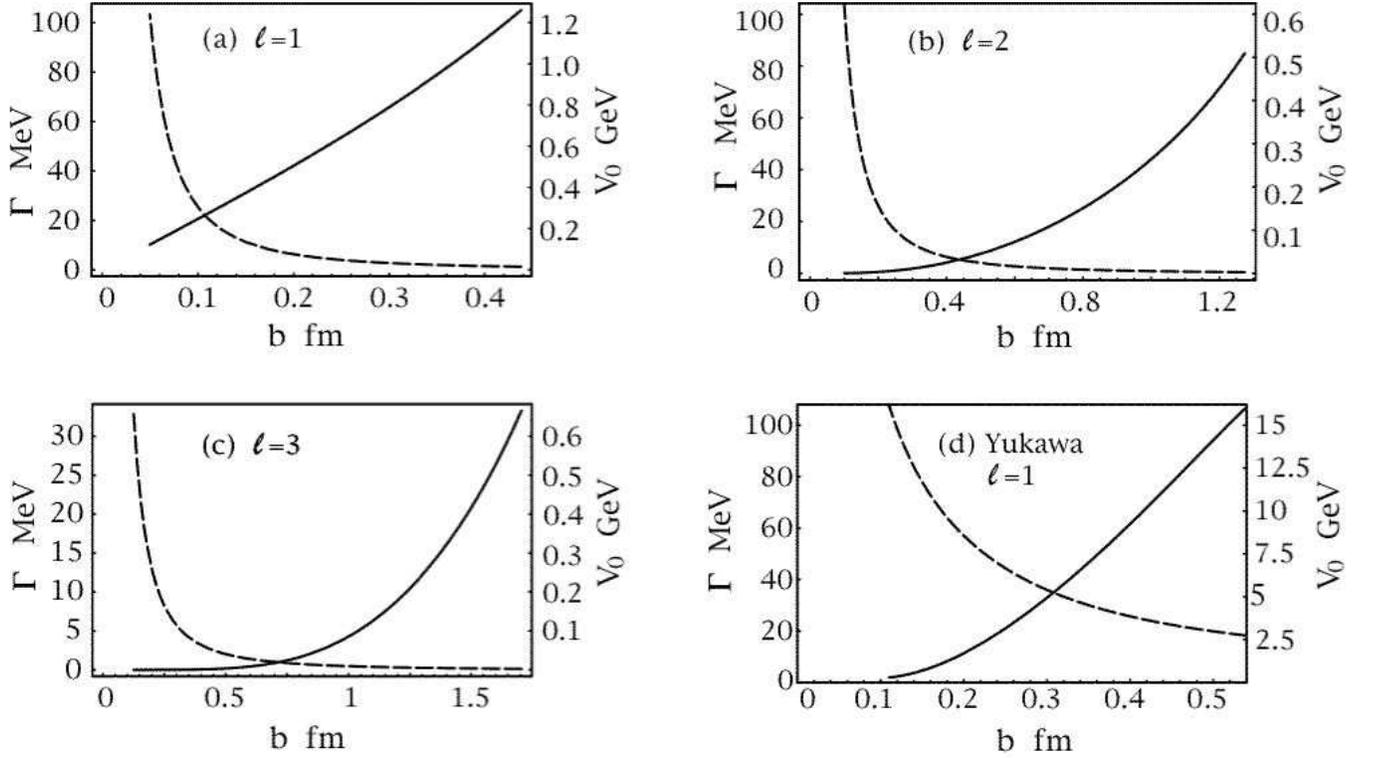}
\caption{\label{figure1} The width of the $\Theta$
in MeV as a function of the range of the attractive potential in
Fermis with the mass of the $\Theta$ held fixed at 1540 MeV, for a
square well, with (a) $\ell=1$; (b) $\ell=2$; and (c) $\ell=3$; and for
a Yukawa potential and $\ell=1$, (d).  Also shown is the depth of the potential
(dashed). }
\end{figure} 

\section*{3.  Feshbach Resonances in a Two Channel Model}

Having failed to find a narrow $\Theta$ in a low partial wave in $KN$
potential scattering, we explore the possibility of adding the
resonance by hand as a bound state in the $KN$ continuum which can
decay to $KN$.  We review the formulation and solution of this simple
scattering problem since it is not so widely known.  

In addition to the open $KN$ channel (of definite isospin) we
introduce a second channel which only has discrete eigenstates in the
energy range of interest.  The two channels are coupled by an
off-diagonal potential, $V_{12}(\vec x)$, which without loss of
generality we take to be real.  We do not need to know the potential
in channel-2.  The potential in channel-1, $V_{11}(\vec x)$, is also
unimportant.  Since it does not generate the resonance, it merely
introduces a smooth background phase in the neighborhood of the narrow
(by assumption) resonance generated by the bound state in channel-2. 
Therefore we set $V_{11}=0$.  The Hamiltonian for this system is given
by
\begin{equation} 
\mathcal{H}=
\begin{pmatrix}
-\nabla^2 & V_{12}(\vec{x}) \\
V_{12}(\vec{x})& H_{22}
\end{pmatrix}
\label{eq1}
\end{equation}
As explained, we assume that $H_{22}$ has a bound state solution given
by
\begin{equation} \label{eq2}
H_{22}\phi ={k_0}^2\phi,
\end{equation}
with ${k_0} =\sqrt{2 \mu E_0}$= 270 MeV. We want to solve coupled set
of Schr\"odinger equations $\mathcal{H}\Psi=k^2\Psi$ with $\Psi=\bigl(
\begin{smallmatrix} \psi_1 \\ \psi_2 \end{smallmatrix} \bigr)$,
\begin{eqnarray}
	-\nabla^{2}\psi_{1}(\vec x) + V_{12}(\vec x)\psi_{2}(\vec x)
	&=&k^{2}\psi_{1}(\vec x)\nonumber\\
	H_{22}\psi_{2}(\vec x) +V_{12}(\vec x)\psi_{1}(\vec x)
	&=& k^{2}\psi_{2}(\vec x)
	\label{eq1a}
\end{eqnarray}
to obtain the scattering amplitude in channel-1.
The particular solution for the second equation can be written in 
terms of the Green's function in channel-2,
\begin{equation}
\psi_2(\vec x) = -\int
G_{22}(\vec{x},\vec{x}\ ';k)V_{12}(\vec{x}\ ')\psi_1(\vec{x}\ ')d\vec{x}\ '
\label{green}
\end{equation}
Note the absence of a solution, $\psi^{(0)}_{2}(\vec x)$, to the homogeneous equation, $\left(H_{22}-k^{2}\right)\psi^{(0)}_{2}(\vec x)=0$, which is excluded because there is no incoming wave in this confined/closed channel.

This problem is simplified dramatically because the resonance is
narrow.  Therefore we are only interested in values of $k$ close to
$k_{0}$.  In that region $G_{22}$ is dominated by the single pole
associated with the state $\phi$,
\begin{equation}
 G_{22}(\vec{x},\vec{x}\ ';k)\approx-
\frac{\phi(\vec{x})\phi(\vec{x}\ ')}{{k_0}^2-k^2} \quad 
\mbox{for}\quad 
k\approx k_{0}
\label{onestate}
\end{equation}
Note $\phi(x)$ can be taken to be real without loss of generality. 
Substituting eq.~(\ref{onestate}), together with eq.~(\ref{green})
into the first of eqs.~(\ref{eq1a}) we obtain an integro-differential
equation for $\psi_1$:
\begin{equation} \label{eq3}
-\nabla^2 \psi_1(\vec{x}) - \int 
d\vec{x}'\frac{\phi(\vec{x})\phi (\vec{x}\ ')V_{12}(\vec{x})V_{12} 
(\vec{x}\ ')}{\kk} 
\psi_1(\vec{x}\ ') = k^2\psi_1(\vec{x}).
\end{equation}
Although the potential in this effective Schr\"odinger equation is
non-local, it is separable, and therefore amenable to an analytic
treatment.  The form of eq.~(\ref{eq3}) explains the physics: if
$V_{12}$ is ``small'' and $k$ is not close to $k_{0}$, then there is
no interaction and $\psi_{1}$ is a free continuum state --- there is
no scattering.  However, no matter how small $V_{12}$ is, as long as
it is not identically zero, there are values of $k$ close to $k_{0}$
for which the potential in eq.~(\ref{eq3}) is arbitrarily strong and
attractive.  Therefore there is always a resonance at $k\approx k_{0}$
and its width is controlled by the magnitude of $V_{12}$.

Conservation of angular momentum requires $V_{12}(\vec x)$ and
$H_{22}$ to be spherically symmetric.  We assume that the bound state
in channel-2 occurs in the $\ell^{\rm th}$ partial wave, in which case
eq.~(\ref{eq3}) reduces to
\begin{equation} \label{eq4}
-u_{\ell}''(r)+\frac{\llll}{r^2}u_{\ell}(r)-\frac{V_{12}(r)v_\ell(r)}{\kk}\int 
dr'V_{12}(r')v_\ell(r')u_{\ell}(r')=k^2 u_{\ell}(r)
\end{equation}
where $\psi_{1}(\vec x) = \frac{1}{r} u_{\ell}(r)Y_{\ell m}(\Omega)$
and $\phi(\vec x)=\frac{1}{r}v_\ell(r)Y_{\ell m}(\Omega)$.  We require
$u_{\ell}(r)/r$ to be regular at the origin.  For $V_{12} \equiv 0$
the solution of the above equation is given by ${\mathcal{U}}_{\ell}
\propto  k r j_{\ell}(kr)$; for non-vanishing $V_{12}$ the solution can
be written,
\begin{equation} \label{eq6}
u_{\ell}^{(+)}(r) = {\mathcal{U}}_{\ell}(r) + \int_0^{\infty} 
dr'G_{\ell}^{(+)}(r,r';k) \frac{V_{12}(r')v_\ell(r')}{\kk} 
\int_0^{\infty} 
dr''V_{12}(r'')v_\ell(r'') u_{\ell}^{(+)}(r''),
\end{equation} 
where we choose the Greens function $G_{\ell}^{(+)}$ to obey outgoing
wave boundary conditions corresponding to scattering,
\begin{equation} \label{eq5}
\begin{split}
G_{\ell}^{(+)} (r,r';k)& = \imath k r r' 
j_{\ell}(kr_<)h_{\ell}^{(1)}(kr_>) \\
&  = \frac{rr'}{\pi}\int_{-\infty}^{\infty} dq~q^2 
\frac{h_{\ell}^{(1)} (qr) j_{\ell}(qr')}{q^2-k^2 - \imath \epsilon}.
\end{split}
\end{equation}

We solve eq.~(\ref{eq6}) by multiplying by $V_{12}(r)v_{\ell}(r)$ and 
integrating over $r$.  Using the Cauchy integral representation for 
the Green's function, we obtain,
\begin{equation}
\eta_{\ell}(k) = \frac{\xi_{\ell}(k)}{1-\frac{X_{\ell}(k^2)}{\kk}}
\end{equation}
where
\begin{eqnarray}
 \eta_{\ell}(k) &=& \int_{0}^{\infty} dr u_{\ell}^{(+)}(r) V_{12}(r) 
v_\ell(r) \nonumber\\
\xi_{\ell}(k) &=& \int_0^{\infty} dr r 
j_{\ell}(kr)V_{12}(r)v_\ell(r)
\end{eqnarray}
and
\begin{equation}
X_{\ell}(k^2) = \frac{2}{\pi}\int\limits_0^{\infty}dq 
\frac{q^2}{q^2-k^2-\imath \epsilon} \vert \xi_{\ell}(q) \vert^2.
\end{equation} 

To extract the scattering amplitude we take the limit $r\to\infty$ in 
eq.~(\ref{eq6}) and compare with the expected form
\begin{equation}
\lim\limits_{r \rightarrow \infty}u_{\ell}^{(+)}(r) = e^{\imath 
\delta_{\ell}(k)} \sin(kr-\frac{\ell \pi}{2} + \delta_{\ell}(k)),
\end{equation}
from which we obtain the following partial wave scattering amplitude 
\begin{equation}
kf_{\ell}(k) \equiv e^{\imath \delta_{\ell}(k)} \sin\delta_{\ell}(k) =
 \frac{{ k} \vert \xi_{\ell}(k) \vert^2}{\kk -
X_{\ell}(k)}.
\end{equation}
This constitutes a complete solution to the problem because 
$\xi_{\ell}(k)$ and $X_{\ell}(k)$ depend only on the known functions 
$V_{12}(r), v_{\ell}(r)$ and the free wavefunction $rj_{\ell}(kr)$.

To leading order in the (small) channel coupling the resonance occurs 
at $k_{0}$ and has width, 
\begin{equation}
\Gamma_{\ell}(k_0) = \frac{k_0}{\mu} \vert \xi_{\ell}(k_0) \vert^2,
\label{gamma}
\end{equation}
$\Gamma_{\ell}$ depends strongly on $k_{0}$ through the dependence 
of  $\xi_{\ell}(k)$ on $j_{\ell}(kr)$. 

The width $\Gamma_{\ell}$ depends only on the product of $V_{12}(r)$
and $v_{\ell}(r)$.  For simplicity we once again take a square well,
$V_{12}(r)=-2 \mu \Omega \theta(b-r)$, and we take $v_\ell(r)/r$ to be
the nodeless $\ell$-wave bound state solution for the infinite square
well of range $b$ and depth adjusted such that the energy of the state
is $k_0^2/2\mu$.  The
results do not depend significantly on these choices because we
normalize the parameters to the known width of the $\Lambda(1520)$.
In this simple case the width of the resonance in $\ell^{\rm th}$
partial wave is given by
\begin{equation}
\Gamma_{\ell}(k_0) = 8 \chi_{\ell}^2 \mu \Omega^2 k_0 b^3 
\frac{j_{\ell}^2(k_0b)}{\bigl [\chi_{\ell}^2 -k_0^2b^2 \bigr ]^2} ,
\label{result}
\end{equation}

\begin{figure}
\includegraphics[width=18cm]{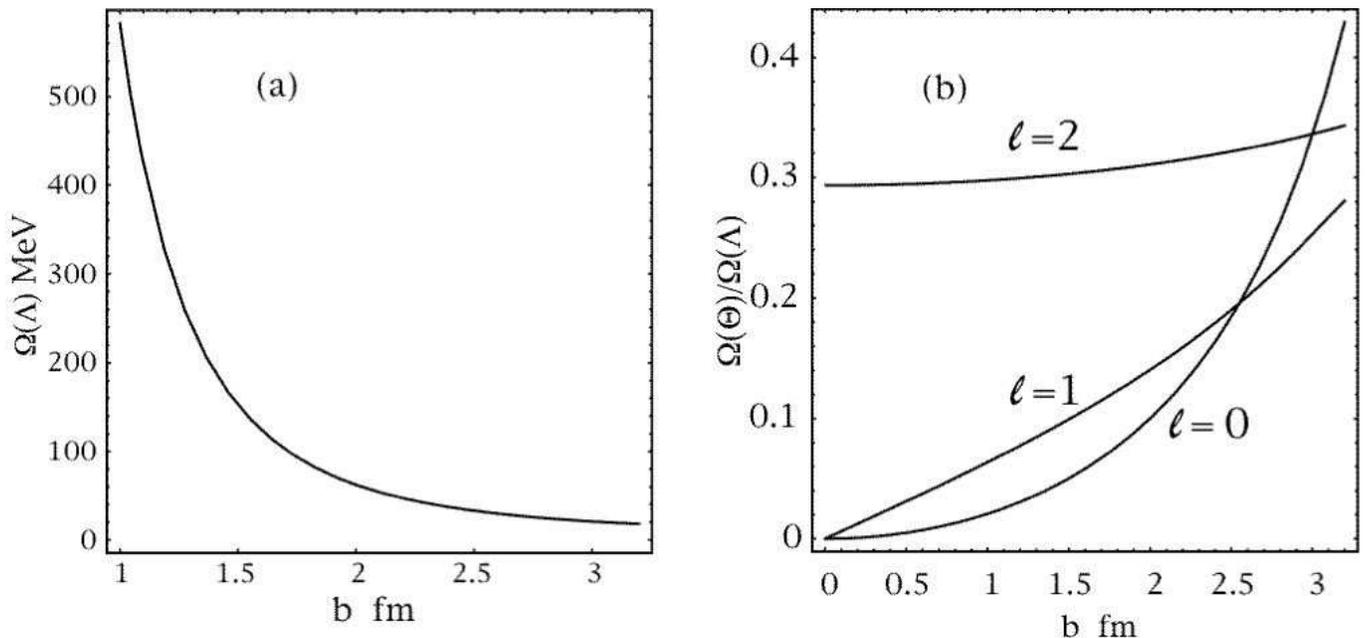}
\caption{\label{figure2} (a) The strength of the
channel coupling potential responsible for the width of the
$\Lambda(1520)$ in the Feshbach-Fano formalism (see
eq.~(\ref{result})).  (b) Ratio of the channel coupling potential
strengths for the $\Theta$ and $\Lambda(1520)$ for various assignments
of the orbital angular momentum in the $\Theta$ channel. }
\end{figure} 

\noindent where $\chi_{\ell}$ is the first zero of $j_{\ell}(x)$.

This result allows us to relate the range, $b$, and the strength,
$\Omega$, of the channel coupling to the width of the resonance as a
function of $\ell$.  Fortunately, as we have noted in the
introduction, Nature has provided us with a conventional non-exotic
$\overline{K}N$ resonance, the $\Lambda(1520)$ at nearly the same mass as the
$\Theta^{+}(1540)$.  The $\Lambda(1520)$ is not a controversial
resonance.  It is usually described as a $p$-wave $uds$ state, and for
the purposes of this analysis we can represent it in $\overline{K}N$ scattering
as a Feshbach resonance.  The analysis we have just completed therefore applies
without alteration to the $\Lambda(1520)$.  It appears in the $d$-wave
and has a partial width into $\overline{K}N$ of 7 MeV, so we have enough
information to use eq.~(\ref{result}) to determine $\Omega(\Lambda)$
as a function of $b$.  This is shown in Fig.~2(a).  This, in turn,
enables us to normalize the values of $\Omega$ needed to account for
the width of the $\Theta^{+}(1540)$.  It is a simple matter to
determine the value of $\Omega(\Theta)$ for some fiducial value of
$\Gamma(\Theta)$ and compare it with $\Omega(\Lambda)$ as a function
of $b$ and $\ell$.  Our results are shown in Fig.~2(b).  There we plot
the ratio $\Omega_{\ell}(\Theta)/\Omega(\Lambda)$ as a function of $b$
for $\ell=0,1$ and 2.  In Fig.~2(b) we have assumed $\Gamma(\Theta)=1$
MeV. When the width of the $\Theta$ is measured one can obtain the
actual value of $\Omega_{\ell}(\Theta)/\Omega(\Lambda)$ by multiplying
the result(s) shown in the figure by $\sqrt{\Gamma(\Theta)}$ in
MeV$^{1/2}$.  The trends in Fig.~2(b) are easy to understand.  As $b$
decreases the channel coupling occurs at distances where
$rj_{\ell}(kb)$ decreases like $kb^{\ell+1}$.  To account for the
width of the ($d$-wave) $\Lambda(1520)$, $\Omega(\Lambda)$ must
increase rapidly as $b\to 0$.  If the $\Theta$ also appears in the
$d$-wave its width scales the same way as the $\Lambda(1520)$'s and
the ratio is only weakly dependent on $b$.  It would be exactly
constant if the masses of the $\Theta$ and the $\Lambda(1520)$ were
equal.  If the $\Theta$ appears in a lower partial wave its width
decreases more slowly as $b\to 0$, so the value of $\Omega(\Theta)$
must decrease relative to $\Omega(\Lambda)$ in order to keep
$\Gamma(\Theta)$ fixed at 1 MeV. Thus as $b\to 0$ the channel coupling
of the $\Theta$ becomes progressively more unnatural.  At $b=1$ fm. 
the ratio $\Omega(\Theta)/\Omega(\Lambda)$ must be 0.02 for $\ell=0$,
0.06 for $\ell=1$, 0.3 for $\ell=2$.  It is in this sense that the a
width of 1 MeV for the $\Theta$ is extraordinary compared to the width
of the conventional $\Lambda(1520)$ resonance.  A theorist attempting
to understand the width of the $\Theta$ implemented as a confined state
appearing in the $KN$ $s$- or $p$-wave, must explain why its coupling
to the open channel is suppressed compared to the coupling of a state
(the $\Lambda(1520)$) with a natural suppression mechanism.  The small
width of the $\Theta$ presents a challenge to model builders even when
it is introduced by hand as a Feshbach resonance.

\section*{Acknowledgements} We would like to thank the members of
the CTP ``Exotica'' working group for many useful discussions.  We also thank the referee for questioning the relation between Feshbach resonances and CDD poles.  This
work is supported in part by the U.S.~Department of Energy (D.O.E.)
under cooperative research agreement~\#DF-FC02-94ER40818.


\begin{thebibliography}{99}
\bibitem{jw}  This case was mentioned in Ref.~\cite{jw1}
\end{thebibliography}
\end{document}